\documentclass[prb,aps,twocolumn,amsmath,amssymb,floatfix,showpacs,
superscriptaddress]{revtex4}

\usepackage[dvips]{graphics}
\usepackage{color}
\definecolor{dred}{rgb}{0,0,0.6}

\begin{document}

\title{Spin Hall effect in a Kagome lattice driven by Rashba spin-orbit 
interaction}

\author{Moumita Dey}

\affiliation{Theoretical Condensed Matter Physics Division, Saha
Institute of Nuclear Physics, Sector-I, Block-AF, Bidhannagar,
Kolkata-700 064, India}

\author{Santanu K. Maiti}

\email{santanu@post.tau.ac.il}

\affiliation{School of Chemistry, Tel Aviv University, Ramat-Aviv,
Tel Aviv-69978, Israel}

\author{S. N. Karmakar}

\affiliation{Theoretical Condensed Matter Physics Division, Saha Institute 
of Nuclear Physics, Sector-I, Block-AF, Bidhannagar, Kolkata-700 064, India}

\begin{abstract}

Using four-terminal Landauer-B\"{u}ttiker formalism and Green's function 
technique, in this present paper, we calculate numerically spin Hall 
conductance (SHC) and longitudinal conductance of a finite size Kagome 
lattice with Rashba spin-orbit (SO) interaction both in presence and 
absence of external magnetic flux in clean limit. In the absence of 
magnetic flux, we observe that depending on the Fermi surface topology 
of the system SHC changes its sign at certain values of Fermi energy. 
Unlike the infinite system (where SHC is a universal constant 
$\pm \frac{e}{8 \pi}$), here SHC depends on the external parameters like 
SO coupling strength, Fermi energy, etc. We show that in the presence of 
any arbitrary magnetic flux, periodicity of the system is lost and the 
features of SHC tends to get reduced because of elastic scattering. But 
again at some typical values of flux ($\phi=\frac{1}{2}$, $\frac{1}{4}$, 
$\frac{3}{4} \ldots$, etc.) the system retains its periodicity depending 
on its size and the features of spin Hall effect (SHE) reappears. Our 
predicted results may be useful in providing a deeper insight into the 
experimental realization of SHE in such geometries.

\end{abstract}


\maketitle

\section{Introduction}

Rapid progress in spin based information processing and storage devicing 
technologies has been metamorphosed into an emerging field called 
`spintronics'~\cite{wolf}, revolutionizing nanotechnology with the 
plethora of concepts which are particularly aimed at the exquisite control 
and manipulation of spin degree of freedom in semiconductor structures 
i.e., using the spin as a career of classical or quantum information. 
Despite being a few decades old topic semiconductor spintronics has 
reignited interest to investigate the role of SO interaction in generating 
pure spin current which is the central theme of this newborn branch of 
condensed matter physics. Originating from the relativistic correction to 
the Schr\"{o}dinger equation, SO interaction provides an all electrical 
way to generate and manipulate spin current in a far precise way rather 
than the usual magnetic field based spin control. Longitudinal flow of 
unpolarized charge current through a sample with SO coupling can induce 
non-equilibrium spin accumulation at the lateral edges of the sample in 
transverse direction, and therefore, a pure spin current is established 
if connected through ideal leads in transverse direction. This is the basic 
phenomenon of Spin Hall Effect. The main source of SO coupling in mesoscopic 
systems comes from either magnetic impurities (extrinsic type) or from 
structural or bulk inversion asymmetry in the confining potential of the 
system (intrinsic type) yielding Rashba or Dresselhaus type SO 
interaction~\cite{rashba,dressel,winkler}.

Few years back some theoretical proposals were made on the existence of 
intrinsic SHE in hole doped~\cite{mura1,mura2} or electron doped~\cite{sinova} 
semi-conducting systems where SO interaction strength is strong enough to 
split the Bloch energy bands for up and down spin electrons. In this case 
a pure spin current is predicted to flow along $Y$ direction, in response 
to the longitudinal electric field along $X$ direction through the 
infinite homogeneous system, essentially capturing the essence of a 
semi-classical effect. Sinova {\em et al.} predicted an universal value of 
Spin Hall conductivity (equal to $\pm \frac{e}{8 \pi}$) for two-dimensional 
electron gas (2DEG) in the clean limit being independent of the SO coupling 
\begin{figure}[ht]
{\centering \resizebox*{8cm}{8cm}{\includegraphics{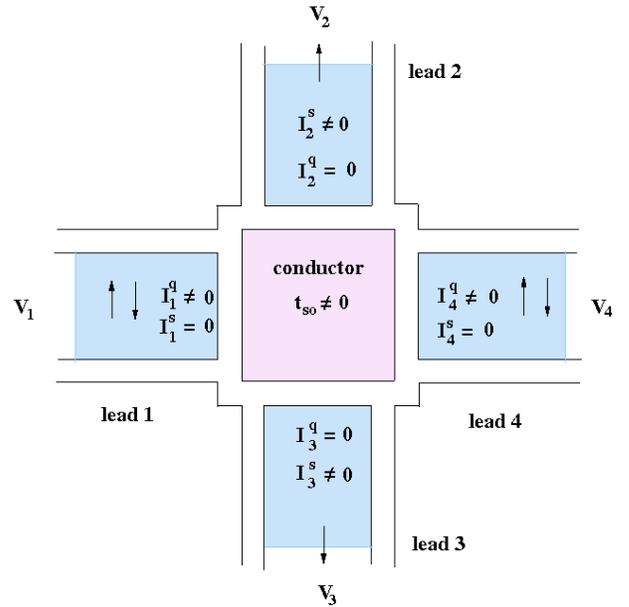}}\par}
\caption{(Color online). Four-probe set-up for measuring spin Hall effect.
An unpolarized charge current is allowed to pass through the longitudinal 
leads, $1$ and $4$, and because of Rashba SO interaction pure spin 
current (no charge current is associated with it, as they are voltage 
probes) flows through the transverse leads, $2$ and $3$.}
\label{spinhall1}
\end{figure}
strength and electron density. The intrinsic SHE is much different from 
the extrinsic effect proposed by Hirsch~\cite{hirsch}, where up and down 
spin electrons get deflected in opposite directions due to spin dependent 
scattering off impurities. However, the magnitude of spin Hall current in 
the intrinsic case is expected to be several orders of magnitude larger 
than the extrinsic one. These theoretical anticipations have also been 
verified experimentally. SHE in $2$-dimensional ($2$D) hole~\cite{wunder} 
or electron~\cite{sih} gases have attracted a tremendous attention in 
research community. Furthermore, in some recent 
experiments~\cite{wunder,kato} the existence of spin accumulation on the 
lateral edge of a two-terminal $2$D hole gas or a $3$D n-type semiconducting 
system has also been detected.

Later on various theoretical investigations have been done to illustrate 
SHE in mesoscopic 2DEG and ring geometries focusing on different aspects 
of this intriguing phenomena. In 2004 Sinova {\em et al.} studied SHE by
measuring dc voltage drop in response to a longitudinal dc
current~\cite{sinova1}. In 2005, Sheng {\em et al.} further investigated 
SHE in a finite size square lattice~\cite{sheng} to explore the 
non-quantized nature of spin Hall conductances which also depends on 
various physical parameters such as SO coupling strength, electronic Fermi 
energy, and disorder strength, etc. In the same year, Nikolic and co-workers 
observed~\cite{niko1} quasi-periodic oscillations in spin Hall
conductances due to spin sensitive quantum interference effect in a 
finite width mesoscopic ring. In another work, again they studied SHE in 
2DEG in detail~\cite{niko2} both for ordered and disordered cases. After 
that, in 2007, Nicolik {\em et al.} made a comparative study between 
extrinsic and intrinsic SHEs in disordered mesoscopic multi-terminal 
systems, again considering a square lattice topology~\cite{niko3}.

Till date a wealth of literature has been formed studying SHE. But the focus
of most of those works were on p- or n-doped semiconductors. Even the 
numerical calculations for finite size systems were based on either square 
lattice~\cite{li,moca1} or ring~\cite{moca2} geometries. Later, discovery 
of a new class of spin Hall insulators~\cite{mura3} by Murakami {\em et al.} 
establishes the fact that apart from the SO interaction strength the lattice 
structure itself plays a very significant role in determining spin Hall 
conductances through its band structure. In 2009, Liu {\em et al.} observed 
more exotic SHE in Kagome~\cite{kagome} and Honeycomb~\cite{honey} lattices 
because of their fascinating topology. But their analysis were based on 
Kubo formalism~\cite{sinova}, which requires an infinite homogeneous system 
in clean limit. Hence a deeper insight into the experimental detection of 
such effect in this kind of geometry demands a quantitative prediction of 
spin Hall conductances in finite size systems. This is the main motivation 
behind our work. 

This kind of geometry is interesting not only from the 
theoretical point of view, but it has also profound experimental 
significance~\cite{expt1,expt2,expt3}. This type of lattice can be easily 
fabricated by modern patterning technique~\cite{pattern}, or observed in 
reconstructed semiconductor surfaces~\cite{recons}. The presence of a 
transverse magnetic field has a deeper impact on electronic properties 
in a kagome lattice which has also been reflected in our SHE study. In 1994, 
Nori {\em et al.} have studied quantum interference effect due to electronic 
motion on kagome lattice in presence of a perpendicular magnetic 
field~\cite{nori1}. Later, in 2002, they have investigated both analytically 
and numerically the mean field superconducting-normal phase boundaries in 
several $2$D networks considering a transverse magnetic field. They 
have determined the transition temperature as a function of magnetic flux
$\phi$ passing through the smallest triangular plaquettes using the 
lattice path integral technique~\cite{nori6}.

In the present work we explore different aspects of SHE in a Kagome 
lattice geometry with Rashba type SO interaction in clean limit using 
Landauer-B\"{u}ttiker formalism. To the best of our knowledge no such 
finite size spin Hall conductance calculation has been done for a kagome 
lattice geometry so far.

The paper is organized as follows. After presenting a brief introduction 
and motivation in Section I, in Section II, we describe the model and 
theoretical formulation to obtain the SHC and longitudinal conductance. 
The numerical results are illustrated in Section III. Finally, in Section 
IV, we summarize our results.

\section{Theoretical formulation}

\subsection{Model and Hamiltonian}

We start describing our model which corroborates with the experimental 
set-up for observing the SHE, where a four-probe mesoscopic bridge is used 
for detection of pure spin current. Four ideal leads are attached to the 
central region (see Fig.~\ref{spinhall1}) which is a finite size kagome 
lattice with Rashba type SO interaction. An unpolarized charge current 
is allowed to pass through the longitudinal leads (lead-$1$ and lead-$4$) 
inducing spin Hall current in the transverse direction (lead-$2$ and 
lead-$3$).

A discrete lattice model is used to describe the finite size kagome lattice 
and also the side attached leads within the framework of tight-binding 
approximation assuming only nearest-neighbor coupling. The Hamiltonian 
representing the entire system can be written as a sum of three terms,
\begin{equation}
H= H_{kag} + H_{leads} + H_{tun}.
\label{eqn1}
\end{equation}
The first term represents the Hamiltonian for the finite size kagome lattice
and it reads as,
\begin{equation}
H_{kag} = \sum_i \mbox{\boldmath $c$}_{i}^{\dag} 
\mbox{\boldmath $\epsilon$} \mbox{\boldmath $c$}_i  
+ \sum_{\langle ij \rangle}  \mbox{\boldmath $c$}_{i}^{\dag} 
\mbox{\boldmath $\tilde{t}$}_{ij} \mbox{\boldmath $c$}_{j} 
+ \sum_{\langle ij \rangle} i\,t_{so} \mbox{\boldmath $c$}_{i}^{\dag} 
[\vec{\sigma} \times \hat{d}_{ij}]_z \mbox{\boldmath $c$}_{j}
\label{eqn2}
\end{equation}
where, \\
$\mbox{\boldmath $c$}^{\dagger}_{i}=\left(\begin{array}{cc}
c_{i,\uparrow}^{\dagger} & c_{i,\downarrow}^{\dagger} 
\end{array}\right);$
$\mbox{\boldmath $c$}_{i}=\left(\begin{array}{c}
c_{i,\uparrow} \\
c_{i,\downarrow}\end{array}\right);$
$\mbox{\boldmath $\epsilon$}=\left(\begin{array}{cc}
\epsilon & 0 \\
0 & \epsilon \end{array}\right)$ \mbox{and} \\
$\mbox{\boldmath $\tilde{t}$}_{ij}=\tilde{t}_{ij}\left(\begin{array}{cc}
1 & 0 \\
0 & 1 \end{array}\right).$\\
~\\
Here, $\epsilon$ is the on-site potential energy and for a perfectly 
ordered system it is set equal to $0$ for all the atomic sites. 
$c_{i,\sigma}^{\dag}$ and $c_{i, \sigma}$ correspond to the creation and 
annihilation operators, respectively, of an electron with spin $\sigma$ 
at the $i$-th site of the conductor. $\tilde{t}_{ij}$ represents the 
isotropic hopping strength between nearest-neighbor sites in presence 
of magnetic field. The effect of the magnetic field $\vec{B}$ 
($= \vec{\nabla} \times \vec{A}$) is incorporated in the hopping term 
$\tilde{t}_{ij}$ through the Peierl's phase factor and it can be written as,
\begin{equation}
\tilde{t}_{ij} = te^{-i\frac{2\pi}{\phi_0} \int 
\limits^{\vec{r}_j}_{\vec{r}_i} \vec{A}.\vec{dl}} \nonumber
\end{equation}
where $t$ is the hopping strength in absence of magnetic field. 
$\phi_0 (=hc/e)$ is the elementary flux quantum. The specific choice of 
the vector potential $\vec{A}$ in this case and the exact calculation of 
the Peierl's phase factor are discussed in the subsequent sub-section. 
\begin{figure}[ht]
{\centering \resizebox*{8cm}{6.5cm}{\includegraphics{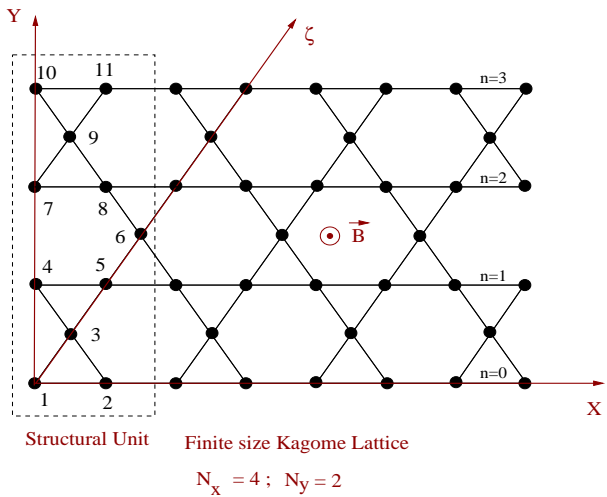}}\par}
\caption{(Color online). Schematic view of a Kagome lattice in presence
of a perpendicular magnetic field where the structural unit (dashed region) 
and the co-ordinate axes are shown. The length and width of the lattice strip 
are determined by two parameters, viz, $N_x$ and $N_y$. Here, $N_x$ represents 
the number of structural units whereas $N_y$ is expressed as $(n_{max}+1)/2$, 
where $n_{max}$ represents the total number of horizontal lines parallel to 
$X$ axis. For this geometry $n_{max}=3$.}
\label{kagome1}
\end{figure}
The third term corresponds to the Rashba type SO coupling in the system 
through spin dependent nearest-neighbor hopping term which introduces 
spin flipping in the system. The quantity $t_{so}$ estimates the strength 
of the Rashba SO interaction originated due to structural inversion asymmetry 
of the confining potential and different band offsets at the heterostructure 
quantum well interface. In this term, $\vec{\sigma}$ is the spin angular 
momentum of the electron and $\hat{d}_{ij}$ is the unit vector along the 
direction from $i$-th site to $j$-th site.

The four metallic leads attached to the conductor are considered to be 
semi-infinite and ideal, i.e., without any disorder and SO interaction.
The leads are described by a similar non-interacting single particle 
Hamiltonian as written below.
\begin{equation}
H_{leads} = \sum_{\alpha=1,2,3,4} H_{\alpha} 
\label{eqn3}
\end{equation}
where,
\begin{equation}
H_{\alpha} = \sum \limits_{n}\epsilon_{l} c_{n}^{\dag}c_n  + 
\sum_{\langle mn \rangle} t_l c^{\dag}_{m} c_n.
\label{eqn4}
\end{equation}
Similarly, the conductor-to-lead coupling is described by the following 
Hamiltonian.
\begin{equation}
H_{tun} = \sum_{\alpha=1,2,3,4} H_{tun,\alpha}
\label{eqn5}
\end{equation}
Here,
\begin{equation}
H_{tun, \alpha} = t_c[c^{\dag}_{i} c_m + c^{\dag}_{m} c_i]
\label{eqn6}
\end{equation}
In the above expression, $\epsilon_l$ and $t_l$ stand for the site energy 
and nearest-neighbor hopping between the sites of the leads. The coupling 
between the leads and the conductor is defined by the hopping integral 
$t_c$. In Eq.~\ref{eqn6}, $i$ and $m$ belong to the boundary sites of the 
kagome ribbon and leads, respectively. The summation over $\alpha$ is due 
to incorporation of the four side-attached leads.

\subsection{Calculation of the Peierl's phase factor}

Now we proceed to evaluate the Peierl's phase factor in the term 
$\tilde{t}_{ij}$.

We choose the vector potential $\vec{A}$ in the form,
\begin{equation}
\vec{A} = -By\:\hat{x} + \frac{By}{\sqrt{3}}\:\hat{y} = 
(-1,\frac{1}{\sqrt{3}},0)\,By.
\label{eqn7}
\end{equation}
This specific choice is followed from a literature~\cite{schreiber}, and 
the purpose of doing that is solely the simplification of the factor 
$\int \vec{A}.\vec{dl}$ along a particular direction ($\zeta$ axis in 
this case).

\begin{figure}[ht]
{\centering \resizebox*{4.5cm}{5cm}{\includegraphics{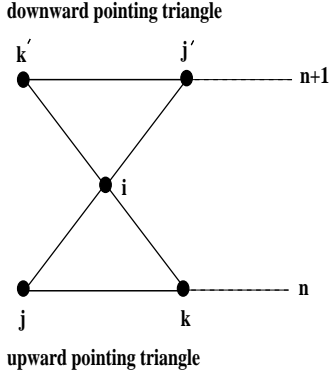}}\par}
\caption{(Color online). Upward and downward pointing triangles labeled
with proper site indices.}
\label{kagome2}
\end{figure}

With this particular choice of $\vec{A}$ we determine $\tilde{t}_{ij}$ for 
three different types of hopping paths in the kagome lattice as follows.
\vskip 0.25cm
\noindent
$\bullet\:\underline{\textbf{Case\;1:}}$
Our choice of gauge ensures that the component of $\vec{A}$ along $\zeta$
(see Fig.~\ref{kagome1}) axis is zero. Therefore, $\tilde{t}_{ij} =t$, 
for an electron moving along $\zeta$ axis ($+$ve, $-$ve or its parallel 
direction).
\vskip 0.25cm
\noindent
$\bullet\:\underline{\textbf{Case\;2:}}$
If we consider the motion along $X$ axis, in general for the $n$-th line
(Fig.~\ref{kagome1}) we can write the hopping integral as,
\begin{eqnarray}
\tilde{t}_{ij} & = & t\,e^{\frac{i 8 \pi n \phi}{\phi_0}} 
\:\mbox{\scriptsize{(hopping\;along\;+ve\;X\;axis)}} \nonumber\\
& = & t\,e^{\frac{-i 8 \pi n \phi}{\phi_0}} 
\:\mbox{\scriptsize{(hopping\;along\;-ve\;X\;axis)}}
\label{eqn8}
\end{eqnarray}
where, $\phi$ is the flux through a smallest triangle of the lattice, 
and $n = 0,1,2,3,\ldots, (2 N_y -1)$.
\vskip 0.25cm
\noindent
$\bullet\:\underline{\textbf{Case\;3:}}$
Finally, we consider the hopping along $k$-th site to $i$-th site and all 
its parallel directions (see Fig.~\ref{kagome2}).

It can be shown by straightforward algebra that for an upward pointing 
triangle ($\bigtriangleup ijk$) the modified hopping strengths are given 
by,
\begin{eqnarray}
\tilde{t}_{k \rightarrow i} = t\,e^{-\frac{i 8 \pi \phi}
{\phi_0}(n+\frac{1}{4})} \hskip 0.2cm \mbox{and} \hskip 0.2cm
\tilde{t}_{i \rightarrow k} = t\,e^{\frac{i 8 \pi \phi}{\phi_0}
(n+\frac{1}{4})}
\label{eqn9}
\end{eqnarray}
The value of $n$ belongs to the base line of the triangle.\\

Similarly, for a downward pointing triangle 
($\bigtriangleup ik^{\prime}j^{\prime}$) the modified hopping integrals 
can be written as,
\begin{eqnarray}
\tilde{t}_{i \rightarrow k^{\prime}} = te^{-\frac{i 8 \pi \phi}
{\phi_0}(n+\frac{3}{4})}\hskip 0.15cm \mbox{and} \hskip 0.15cm
\tilde{t}_{k^{\prime} \rightarrow i} & = & te^{\frac{i 8 \pi \phi}
{\phi_0}(n^{\prime}-\frac{1}{4})}
\label{10}
\end{eqnarray}
Here, $n^{\prime} = n+ 1$, as for a downward pointing triangle the 
sites $j^{\prime}$ and $k^{\prime}$ does not belong to the same 
value of $n$.

Following the above prescription we incorporate the 
effect of magnetic field quite easily in our lattice geometry. But there 
are other ways through which we can introduce the effect of magnetic 
field in lattice models. For example, Nori {\em et al.} have 
used a different approach of lattice path integral technique~\cite{nori1,
nori6,nori2,nori3,nori4,nori5} to describe the effect of magnetic field 
in different lattice structures. Particularly, this technique allows us 
to evaluate the physical quantities in terms of explicit functions for a 
continuously tunable flux.

\subsection{Expressions of Longitudinal ($G_L$) and spin Hall conductances 
($G_{SH}$)}

According to spin Hall phenomenology, in our model pure spin current is 
predicted to flow through the transverse leads, due to the flow of charge
current through the longitudinal leads. Hence, the linear response 
longitudinal (4-probe) and Spin Hall conductances are defined as
\begin{equation}
G_L=\frac{I^{q}_{4}}{V_1-V_4}
\label{eqn11}
\end{equation}
and
\begin{equation}
G_{sH}=\frac{\hbar}{2e}\frac{I^{s}_{2}}{V_1-V_4}
\label{eqn12}
\end{equation}
where, $I_{4}^{q}$ and $I_{2}^{s}$ are the charge current and spin current
flowing through the lead-4 and lead-2, respectively. $V_{m}$ $(m=1,2,3$ 
and $4)$ is the potential at the $m$-th lead.

Now, following Landauer-B\"{u}ttiker multi-probe formalism the charge and 
spin currents flowing through the lead $m$ with potential $V_m$, can be 
written in terms of spin resolved transmission probabilities as~\cite{pareek},
\begin{equation}
I_{m}^{q}=\frac{e^2}{h}\sum_{n,\sigma^{\prime},\sigma} 
(T_{n m}^{\sigma \sigma^{\prime}}V_m - T_{m n}^{\sigma \sigma^{\prime}}V_n)
\label{eqn13}
\end{equation}
\begin{equation}
I_{m}^{s}=\frac{e^2}{h}\sum_{n,\sigma^{\prime}} \left[ 
(T^{\sigma^{\prime} \sigma}_{n m} - T^{\sigma^{\prime} -\sigma}_{n m})V_m
+(T^{-\sigma \sigma^{\prime}}_{m n} - T^{\sigma \sigma^{\prime}}_{m n})
V_n\right]
\label{eqn14}
\end{equation}
Considering linear transport regime, at absolute zero temperature the
linear conductance $(G_{pq})$ is obtained using Landauer conductance 
formula~\cite{land},
\begin{equation}
G_{pq}^{\sigma \sigma^{\prime}} =\frac{e^2}{h}T_{pq}^{\sigma 
\sigma^{\prime}}(E_F).
\label{eqn15}
\end{equation}
Using Landauer conductance formula spin current through $m$-th lead can 
be re-written in terms of spin resolved conductances as,
\begin{eqnarray}
I^{s}_{m} & = & \sum_{n} [(G_{nm}^{\uparrow \uparrow} + 
G_{nm}^{\downarrow \uparrow} - G_{nm}^{\uparrow \downarrow} 
- G_{nm}^{\downarrow \downarrow})V_m \nonumber\\
& & + (G_{mn}^{\downarrow \uparrow} + 
G_{mn}^{\downarrow \downarrow} - G_{mn}^{\uparrow \uparrow} 
- G_{mn}^{\uparrow \downarrow})V_n ]
\label{eqn16}
\end{eqnarray}
Equation~\ref{eqn16} can be simplified in terms of two quantities defined 
as follows.
\begin{eqnarray}
G^{in}_{mn} & = & G_{mn}^{\uparrow \uparrow} + G_{mn}^{\uparrow \downarrow}
 - G_{mn}^{\downarrow \uparrow} - G_{mn}^{\downarrow \downarrow} \nonumber\\
G^{out}_{mn} & = & G_{mn}^{\uparrow \uparrow} + G_{mn}^{\downarrow \uparrow}
 - G_{mn}^{\uparrow \downarrow} - G_{mn}^{\downarrow \downarrow}
\label{eqn17}
\end{eqnarray}
Physically the term $\sum_n G_{nm}^{out} V_m$ is the total spin current 
flowing from the $m$-th lead with voltage $V_m$ to all other $n$ leads,
while the term $\sum_n G_{mn}^{in} V_n$ defines the total spin current 
flowing into the $m$-th lead from the all other $n$ leads having potential
$V_n$.

Therefore, the spin current through lead $m$ becomes, 
\begin{equation}
I_{m}^{s} = \sum_{n} \left[ G_{nm}^{out} V_m -G_{mn}^{in} V_n \right]
\label{eqn18}
\end{equation}
Hence, following spin Hall phenomenology, in our set-up since the transverse 
leads are voltage probes, the net charge currents through lead-$2$ and 
lead-$3$ are zero i.e., $I_2^q=I_3^q=0$. On the other hand, as the currents
in the various leads depend only on voltage differences among them, we can 
set one of the voltages to zero without any loss of generality. Here,
we set $V_4=0$. So, from Eq.~\ref{eqn18} we have,
\begin{equation}
I^{s}_{2} = \left(G_{12}^{out} + G_{32}^{out} + G_{42}^{out}\right)V_2 
- G_{23}^{in}V_1 - G_{21}^{in}V_1
\label{eqn19}
\end{equation}
Hence, the expression of spin Hall conductance becomes 
\begin{equation}
G_{sH} = \frac{\hbar}{2 e} \left[ (G_{12}^{out} + G_{32}^{out} 
+ G_{42}^{out}) \frac{V_2}{V_1} - G_{23}^{in}\frac{V_3}{V_1} 
- G_{21}^{in}  \right]
\label{eqn20}
\end{equation}
This is the most general expression of SHC, but it can further be 
simplified if we assume that the leads are connected to a geometrically 
symmetric ordered bridge, so, $\frac{V_3}{V_1} = \frac{V_2}{V_1} = 0.5$.
Now, in absence of external magnetic flux only the Rashba SO interaction 
does not break the time reversal symmetry (TRS). In general, for a time 
reversal invariant system $G_{pq}^{\sigma \sigma^{\prime}} = 
G_{qp}^{- \sigma^{\prime} -\sigma}$, which is equivalent to write
$G_{pq}^{in} = - G_{qp}^{out}$.
\noindent
Therefore, obeying TRS the expression for the spin Hall conductance can 
be written in terms of spin resolved transmission probabilities in a 
compact form as,
\begin{equation}
G_{sH} = \frac{e}{8 \pi} \left[T_{42}^{out} + 2\;T_{32}^{out} 
+ 3\;T_{12}^{out}  \right]
\label{eqn21}
\end{equation}
One important point to be noted here is that unlike the charge current 
spin current is a vector quantity, which immediately gives rise to the 
three different components of spin Hall conductances ($G_{sH}^{x}$,
$G_{sH}^{y}$ and $G_{sH}^{z}$) and they are defined as follows.
\begin{eqnarray}
G_{sH}^{x}=I_2^x/(V_1-V_4) \nonumber\\
G_{sH}^{y}=I_2^y/(V_1-V_4) \nonumber\\
G_{sH}^{z}=I_2^z/(V_1-V_4)
\label{eqn21new}
\end{eqnarray}
Using Eq.~\ref{eqn21} all the three different components of SHC can be 
evaluated, only the choice of basis while constructing the matrices is 
important. For example, if we are working in $\sigma_z$ diagonal 
representation (i.e., the $Z$ axis chosen to be the spin quantization 
axis), then Eq.~\ref{eqn21} gives the $z$-component of SHC. Similarly,
other components can also be evaluated by a simple unitary transformation 
to the basis set. In the present work we are working only with the 
$z$-component of SHC.

In a similar way from Eqs.~$\ref{eqn11}$ and $\ref{eqn13}$ the expression
of longitudinal conductance can be written as
\begin{equation}
G_L = \frac{e^2}{h}\left[T_{41}+ 0.5 \;T_{42} + 0.5 \;T_{43}\right].
\label{eqn22}
\end{equation}

\subsection{Evaluation of the Transmission Probability by Green's function 
technique}

To obtain the transmission probability of an electron through such a
four-probe mesoscopic bridge system, we use Green's function formalism. 
Within the regime of coherent transport and in the absence of Coulomb 
interaction this technique is well applied.

The single particle Green's function operator representing the entire
system for an electron with energy $E$ is defined as,
\begin{equation}
G=\left(E - H + i\eta \right)^{-1}
\label{eqn23}
\end{equation}
where, $\eta \rightarrow 0^+$.

Following the matrix forms of \mbox{\boldmath $H$} and \mbox{\boldmath $G$}
the problem of finding \mbox{\boldmath $G$} in the full Hilbert space of
\mbox{\boldmath $H$} can be mapped exactly to a Green's function
\mbox{\boldmath $G_{kag}^{eff}$} corresponding to an effective
Hamiltonian in the reduced Hilbert space of the conductor (i.e.,
the kagome lattice itself) and we have,
\begin{equation}
\mbox{\boldmath ${\mathcal G}$=$G_{kag}^{eff}$}=\left(\mbox{\boldmath $E- 
H_{kag}-\sum \limits_{\alpha, \sigma}\Sigma_{\alpha}^{\sigma}$}\right)^{-1}
\label{equ24}
\end{equation}
where,
\begin{equation}
\mbox{\boldmath $\Sigma_{\alpha}^{\sigma}$} = \mbox{\boldmath 
$H_{tun, \alpha}^{\dag} G_{\alpha} H_{tun, \alpha}$} 
\label{eqn25}
\end{equation}
These \mbox{\boldmath $\Sigma_{\alpha}$} ($\alpha=1,2,3$ and $4$) are the 
contact self-energies introduced to incorporate the effect of coupling
of the conductor to the attached ideal leads. It is evident from 
Eq.~\ref{eqn25} that the form of the self-energies are independent of 
the conductor itself through which spin transmission is studied.

Following Lee and Fisher's expression for the probability of an electron 
to transmit from lead q with spin $\sigma^{\prime}$ to lead p with spin 
$\sigma$ can be written as~\cite{lee},
\begin{equation}
T_{pq}^{\sigma \sigma^{\prime}} = \mbox{Tr}\mbox{\boldmath 
[$\Gamma_{p}^{\sigma} \mathcal {G}^r \Gamma_{q}^{\sigma^{\prime}} 
\mathcal {G}^a$]}.
\label{eqn26}
\end{equation}
\mbox{\boldmath$\Gamma_{k}^{\sigma}$}'s are the coupling matrices 
representing the coupling between the kagome lattice and the leads, 
and they are mathematically defined by the relation,
\begin{equation}
\mbox {\boldmath $\Gamma_k^{\sigma}$} = i \left[\mbox 
{\boldmath $\Sigma_{k}^{\sigma} - \Sigma_{k}^{\sigma \dag}$}\right]
\label{eqn27}
\end{equation}
Here, \mbox{\boldmath $\Sigma_{k}^{\sigma}$} and 
\mbox{\boldmath $\Sigma_{k}^{\sigma \dag}$}
are the retarded and advanced self-energies associated with the $k$-th
lead, respectively.

It is shown in literature by Datta {\em et al.}~\cite{datta1,datta2} 
that the self-energy can be expressed as,
\begin{equation}
\mbox{\boldmath ${\Sigma^{\sigma}_{k}}$} = \mbox{\boldmath 
$\Lambda_{k}^{\sigma}$} - i \mbox{\boldmath $\Delta_{k}^{\sigma}$}.
\label{eqn28}
\end{equation}
The real part of self-energy describes the shift of the energy levels
and the imaginary part corresponds to the broadening of the levels. The
finite imaginary part appears due to incorporation of the semi-infinite
leads having continuous energy spectrum. Therefore, the coupling matrices
can easily be obtained from the self-energy expression and is 
expressed in the form,
\begin{equation}
\mbox{\boldmath $\Gamma_{k}^{\sigma}$}=-2\,{\mbox {Im}} 
(\mbox{\boldmath $\Sigma_{k}^{\sigma}$}).
\label{eqn29}
\end{equation}

\subsection{Evaluation of the Self-Energy}

Finally, it remains the evaluation of the self-energies for the finite-width, 
multi-channel, square lattice leads. Now for the semi-infinite longitudinal 
leads (leads $1$ and $4$) as the translational invariance is preserved in 
$X$ direction only, the wave function amplitude at any arbitrary site $m$ 
of the leads can be written as, $\phi_m \propto e^{i k_x m_x a} 
\sin(k_y m_y a)$, 
with energy 
\begin{equation}
E = 2t_L[\cos(k_x a)+ \cos(k_y a)]
\label{eqn30}
\end{equation}

In Eq.~\ref{eqn30}, $k_x$ is continuous, while $k_y$ has discrete values 
given by,
\begin{equation}
k_y(i) = \frac{i \pi}{(M + 1)a}
\label{eqn31}
\end{equation}
Here, $i= 1,2,3 \ldots M$. $M$ is the total number of transverse channels
in the leads, and in our case $M = 2 N_y$.

The self-energy matrices are constructed in the reduced Hilbert space of the 
conductor itself. These matrices have non-zero elements only for the sites 
on the edge layer of the sample coupled to the leads and it is given by,
\begin{equation}
\Sigma^r_{1(4)}(m,n) = \frac{2}{M + 1} \sum_{k_y} \sin(k_y m_y a) 
\Sigma^r (k_y) \sin(k_y n_y a)
\label{eqn32}
\end{equation}
$\Sigma^{r}(k_y)$ is the self-energy of each transverse channel and for a 
specific value of $k_y$ it becomes,
\begin{equation}
\Sigma^r (k_y) = \frac{t_c^2}{2t_L^2} \left[E-\epsilon(k_y) -i 
\sqrt{4t_L^2-(E-\epsilon(k_y))^2} \right]
\label{eqn33}
\end{equation}
with $\epsilon(k_y) = 2t_L \cos(k_y a)$, when the energy lies within the 
band, i.e., $|E-\epsilon(k_y)| < 2 t_L$;
and
\begin{equation}
\Sigma^r (k_y) = \frac{t_c^2}{2t_L^2} \left[E-\epsilon(k_y) \mp
\sqrt{(E-\epsilon(k_y))^2-4t_L^2} \right]
\label{eqn34}
\end{equation}
when the energy lies outside the band. Here the $-ve$ sign appears 
for $E > \epsilon(k_y) + 2|t_L|$ and $+ve$ sign comes for 
$E < \epsilon(k_y) - 2|t_L|$.

The self-energy matrices for the other two leads (leads $2$ and $3$) are 
also constructed in a similar way. The only difference in this case is 
that the translational invariance is preserved along $Y$ direction, and 
accordingly, the finite values of $k_x$ are chosen.

\section{Results and discussion}

We start analyzing the numerical results by referring to the values of 
different parameters used for our calculation. Throughout the presentation
we set $\epsilon = \epsilon_{l} =0$, and fix all the hopping integrals 
($t$, $t_l$ and $t_c$) at the value $1$. The energy scale is measured
in unit of $t$ and choose the unit where $c=h=e=1$. The Rashba coupling 
strength $t_{so}$ is also scaled in unit of $t$, and it is usually 
chosen as $t_{so} \lesssim t$. The magnetic flux $\phi$ is measured in 
unit of elementary flux quantum $\phi_0$ (=$hc/e$).

\subsection{Variation of spin Hall conductances with Fermi energy}

\begin{figure}[ht]
{\centering \resizebox*{7.0cm}{8.0cm}{\includegraphics{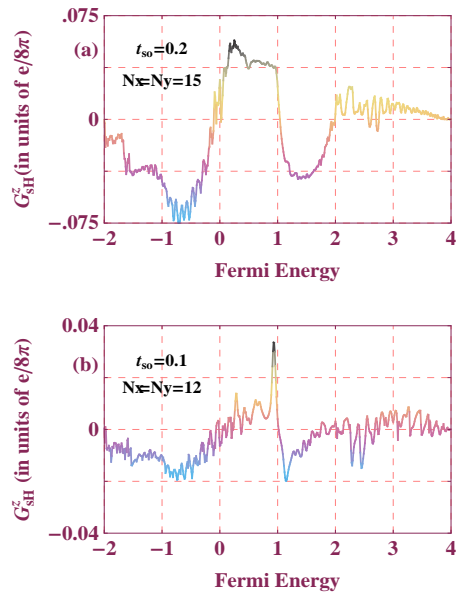}}\par}
\caption{(Color online). Variation of $G_{sH}^{z}$ with Fermi energy in
the absence of external magnetic flux for two different sets of parameter 
values (shown inside the figure).}
\label{hallcond1}
\end{figure}

In Fig.~\ref{hallcond1} we plot the $z$-component of spin Hall conductance 
$(G_{sH}^{z})$ as a function of Fermi energy $(E_F)$ for two different 
values of Rashba coupling strengths and different 
system sizes. This figure demonstrates the fact that unlike the bulk or 
infinite system, in this case the spin Hall conductance does not have a 
universal value ($\pm \frac{e}{8 \pi}$ as predicted in the case of an infinite 
2DEG), rather it depends explicitly on the system parameters like Fermi 
energy ($E_F$), strength of Rashba SO interaction ($t_{so}$), system size, 
etc. The non-zero spin Hall conductance observed in Fig.~\ref{hallcond1} 
is a consequence of the fact that in presence of the Rashba SO interaction
in the conductor, the up and down spin electrons flow in opposite 
transverse directions even in the absence of external magnetic flux, 
leading to a pure spin current, in response to the flow of charge current 
along the longitudinal direction.

It has already been shown in literature that geometries like kagome 
lattice, graphene flakes, etc., exhibit some unique features of the spin 
Hall conductance due to their fascinating structures. 
Following Kubo formalism, Liu {\em et al.} have analyzed the variation of 
conserved spin Hall conductance with respect to the Fermi energy for an 
infinite kagome lattice~\cite{kagome}. Considering a two-band approximation 
and treating the Rashba SO coupling as perturbation, they have analyzed 
the presence of various spin Hall plateaus (at $\pm \frac{e}{8 \pi}$ and 
$\pm \frac{e}{4 \pi}$) with Berry phase interpretation. It has been shown 
that when the Fermi energy lies within the range $-2<E_F<0$, the contribution 
to spin Hall conductivity due to the conventional part ($\sigma_{xy}^{s0}$) 
is $-1$ (in units of $\frac{e}{8 \pi}$). In this case, by expanding the 
Hamiltonian around the \mbox{\boldmath $\Gamma$} point upto first order 
in the Rashba coefficient $t_{so}$ they have established the similarity of 
the Rashba Hamiltonian with that of a semiconductor 2DEG, and by 
straightforward analytical calculation spin Hall conductivity is obtained 
as $-\frac{e}{8 \pi}$. But with the increase in electron filling, within 
the range $0<E_F<1$, expansion of the Hamiltonian around the {\bf K} point 
(`{\bf K}-valley' Hamiltonian) exhibits a Dirac type spectrum with linear 
dependence of energy on momentum, and accordingly, 
the value of conventional spin Hall conductivity gets the value $2$. 
Furthermore the anti-symmetry at the band center is completely due to 
the particle-hole symmetry of the employed tight-binding Hamiltonian. 
The variation in sign and magnitude of spin Hall conductivity in the 
range ($-2<E_F<1$) is associated with the change in Fermi surface topology 
surrounding the high symmetry \mbox{\boldmath $\Gamma$} and {\bf K} 
Brillouin Zone (BZ) points. The analysis is completely suitable for an 
infinite system. {\em Here, in our work we try to investigate numerically 
whether the essence of these predicted features in a realistic, 
finite size, small scale, mesoscopic system are still present or not.}

In this figure (Fig.~\ref{hallcond1}) we observe that for a particular 
system size with finite Rashba coupling strength, the spin Hall conductance 
\begin{figure}[ht]
{\centering \resizebox*{6.5cm}{4.0cm}{\includegraphics{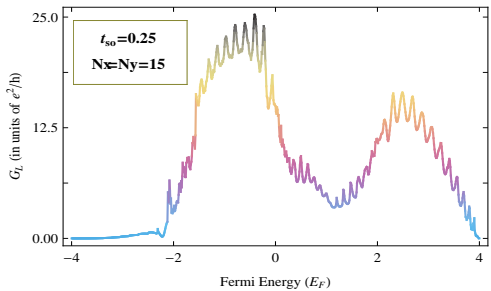}}\par}
\caption{(Color online). Variation of four-probe longitudinal conductance
as a function of Fermi energy in the absence of external magnetic flux.}
\label{longcond}
\end{figure}
is nearly an antisymmetric function with respect to the band center i.e., 
$G^{z}_{sH} = 0$ at $E_F=1$, reflecting the particle-hole symmetry of 
the tight-binding Hamiltonian. It is a consequence of the fact that the 
spin current is defined as the difference between spin resolved charge 
currents ($I_{\uparrow}$ and $I_{\downarrow}$), and the spin current 
carried by negatively charged electrons at $E_F<1$ can be interpreted as 
the propagation of positively charged holes with opposite spin in opposite 
direction. Apart from that, sign reversals take place also at $E_F=0$ and 
at $E_F=2$ as already predicted for the infinite system. The oscillatory 
behavior in SHC for a smaller system size is entirely the finite size effect. 
The feature of sign reversals in SHC gets prominent with the increase of 
system size. Hence, our numerical results for finite sized systems provide 
all the essential features of SHC pattern observed in an infinite kagome 
lattice.

\subsection{Variation of longitudinal conductance with Fermi energy}

In Fig.~\ref{longcond} we explore the variation of four-probe 
longitudinal conductance in presence of finite Rashba interaction strength 
($t_{so} \neq 0$) as a function of Fermi energy. It exhibits quite a similar
behavior to our previous investigation~\cite{kagomeold} of two-terminal 
longitudinal conductance. The addition of the two other semi-infinite 
transverse finite 
width leads allows some extra phase-breaking paths and thereby lifting 
the transmission zeros and hence broadening the conductance peaks. The 
conductance spectrum reveals itself the energy eigenstates of the finite 
size system. The presence of Rashba coupling does not affect the energy 
spectrum in a significant way apart from shifting of the energy eigenvalues 
a little.

\subsection{Spin Hall conductance as a function of Rashba coupling strength}

\begin{figure}[ht]
{\centering \resizebox*{6.5cm}{4.0cm}{\includegraphics{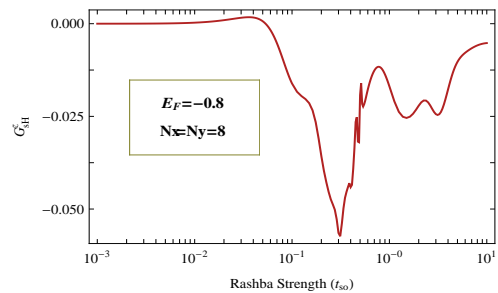}}\par}
\caption{(Color online). Z-component of spin Hall conductance as a function
of Rashba coupling strength in the absence of magnetic flux for a typical 
system size and at a particular Fermi energy.}
\label{hallzrashba}
\end{figure}

The dependence of spin Hall conductance of a finite size kagome lattice on 
the Rashba coupling strength at the typical energy $E_F=-0.8$ is illustrated 
in Fig.~\ref{hallzrashba}. Unlike the results predicted by using linear 
response 
theory, in this case $G_{sH}^z \rightarrow 0$ as $t_{so} \rightarrow 0$, 
and $G_{sH}^{z}$ shows non-negligible, non-zero values within the range 
$0.1 < t_{so} < 5$. The strength of Rashba SO interaction can be tuned 
externally by using a gate voltage. Although the physically accessible 
range of $t_{so}$ (in unit of $t$) is $0.001$ to $1$, but here we plot the 
spectrum for a wider range, a part of which may be beyond the experimental 
reach till date. For a sufficiently large SO coupling, $G_{sH}^{z}$ again 
drops to $0$, because large SO coupling in the conductor forms a large 
potential barrier for the incident electrons, yielding a very small 
probability to transmit through the conductor.

\subsection{Effect of Magnetic field}

\subsubsection{Energy-flux characteristics}

\begin{figure}[ht]
{\centering \resizebox*{7.5cm}{9cm}{\includegraphics{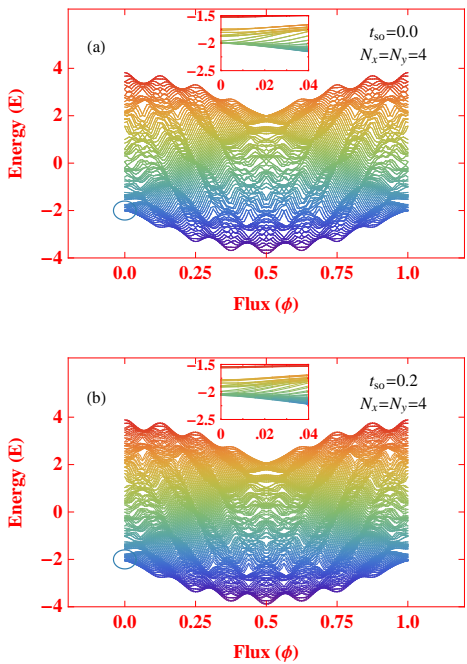}}\par}
\caption{(Color online). Energy-flux characteristics of a finite size 
kagome lattice in (a) absence and (b) presence of Rashba type SO 
coupling. The position of the highly degenerate energy level at $E=-2$
is slightly shifted due to Rashba interaction but the degeneracy is not 
broken.}
\label{energyflux}
\end{figure}

Figure~\ref{energyflux} depicts the energy-flux characteristics of a finite 
size ($N_x=6$ and $N_y=2$) kagome lattice both in the presence and absence of 
Rashba type SO coupling. The energy eigenvalues are obtained by diagonalizing 
the Hamiltonian and the energy-flux spectrum is often called the Hofstader 
spectrum. Form Fig.~\ref{energyflux}(a) it is clearly visible that a highly 
degenerate level exists at $E = -2t$ when $\phi$ is set equal to $0$, and 
the application of a very small non-zero flux starts to break the degeneracy. 
The presence of even a very small magnetic flux affects the phase of the 
electronic wave function and thus tends to destroy the quantum interference 
and eventually breaks the flat band~\cite{kimura}. Again at the half 
flux-quantum ($\phi=\phi_0/2$), the flat band re-appears because of the 
interference effect but at the energy $E=2t$. The energy levels are periodic 
in $\phi$, showing $\phi_0$ (which is set to $1$ in our calculation) 
flux-quantum periodicity and the energy spectrum is mirror symmetric about 
$\phi=0.5$. The presence of a small non-zero SO interaction does not affect 
the degeneracy much which is clearly noticed from Fig.~\ref{energyflux}(b).

\subsubsection{Effect of magnetic flux on spin Hall conductance}

\begin{figure}[ht]
{\centering \resizebox*{7.5cm}{12cm}{\includegraphics{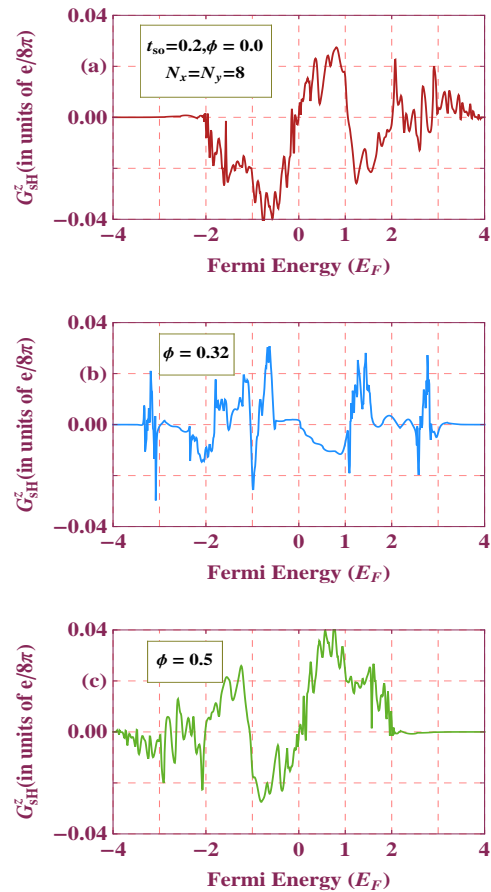}}\par}
\caption{(Color online). Variation of $G_{sH}^{z}$ as a function of Fermi 
energy for three different values of flux, where (a) $\phi=0$ (b) $\phi=0.32$ 
and (c) $\phi=0.5$.}
\label{hallmag}
\end{figure}

In Fig.~\ref{hallmag} we plot the nature of SHC as a function of Fermi 
energy for different values of magnetic flux $\phi$, where (a), (b) and 
(c) correspond to $\phi=0$, $0.32$ and $0.5$, respectively. From the spectra 
it is observed that by applying an arbitrary magnetic flux ($\phi = 0.32$) 
the features of SHE tend to get diminished, but interestingly at 
$\phi=\phi_0/2$, all the above mentioned features re-appear being just the 
inverse of the case for $\phi=0$, in accordance with the change of band 
structure with the application of magnetic flux. These features can be 
illustrated as follows.

\begin{figure}[ht]
{\centering \resizebox*{5.5cm}{5cm}{\includegraphics{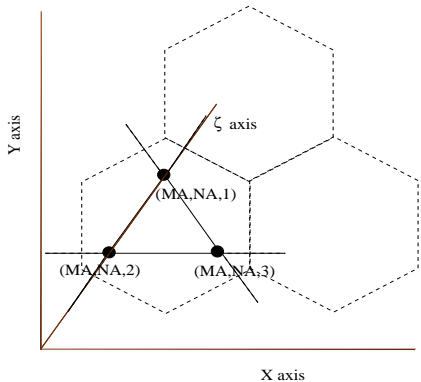}}\par}
\caption{(Color online). Unit cell configuration of an infinite kagome 
lattice.}
\label{unitcell}
\end{figure}

If we write the Schr\"{o}dinger equations for three different sites in a 
unit cell (Fig.~\ref{unitcell}) of an infinite kagome 
lattice~\cite{schreiber}, we get,
\begin{eqnarray}
E\psi(MA,NA,1) & = & \psi(MA,NA,2) + \psi(MA,(N+1)A,2) + \nonumber\\
& & e^{\frac{i 8 \pi \phi}{\phi_0}(N+\frac{1}{4})} \psi(MA,NA,3) + \nonumber\\
& & e^{\frac{-i 8 \pi \phi}{\phi_0}(N+\frac{3}{4})} 
\psi((M-1)A,(N+1)A,3) \nonumber\\
E\psi(MA,NA,2) & = & \psi(MA,NA,1) + \nonumber\\ 
& & \psi((M-1)A,(N-1)A,1)+ \nonumber\\
& & e^{\frac{-i 8 \pi N \phi}{\phi_0}} \psi((M-1)A,NA,3) + \nonumber\\
& & e^{\frac{i 8 \pi N \phi}{\phi_0}}\psi(MA,NA,2) \nonumber\\
E\psi(MA,NA,3) & = & e^{\frac{-i 8 \pi N \phi}{\phi_0}} 
\psi(MA,NA,2)+ \nonumber\\ 
& & e^{\frac{i 8 \pi N \phi}{\phi_0}} \psi((M+1)A,NA,2) + \nonumber\\ 
& &  e^{\frac{-i 8 \pi \phi}{\phi_0}
(N+\frac{1}{4})} \psi(MA,NA,1) + \nonumber\\ 
& & e^{\frac{-i 8 \pi \phi}{\phi_0}(N-\frac{1}{4})}\psi(MA,(N-1)A,1)
\label{eqn35}
\end{eqnarray}
where, $\psi(MA,NA,j)$ denotes the wave amplitude at a particular site 
($MA,NA,j$) of the unit cell. Here, $M$ indicates the $M$-th triangle 
along the $X$ axis, $N$ indicates $N$-th triangle along the $\zeta$ axis 
and $j=1,2$ and $3$ represents the vertex in the triangle.

Here, for any arbitrary flux, translational invariance is lost along
all directions and the hopping integrals become different. As a result 
the sharp features like the sign reversal about the symmetry points of 
the Fermi surface get reduced because of the elastic scattering, which is 
visible in Fig.~\ref{hallmag}(b). Very interestingly, again at 
$\phi=\phi_0/2$,it can be shown from the above equations that the 
translational invariance is again restored along both the $X$ and $\zeta$ 
directions which can be deducted from the above equations, and therefore,
scattering effect gets completely suppressed and SHC re-appears but with 
inverse manner due to the change in band structure.

It can also be observed that for $\phi=\frac{n}{8m}$ ($n$ and $m$ being 
integers), translational invariance is retained (increasing the unit cell 
dimension) and hence system size has an important role as we are dealing 
with a finite size kagome lattice. In this case, some anisotropy is introduced 
in the system through the hopping terms. Therefore, the SHC pattern changes 
accordingly due to the changes in the Fermi surface topology. In our case
($N_x=N_y=8$) we observe that for $\phi = \frac{1}{4}$ and $\frac{3}{4}$ 
a regular pattern in SHC is obtained along with the sign reversal at the 
\begin{figure}[ht]
{\centering \resizebox*{7.5cm}{7cm}{\includegraphics{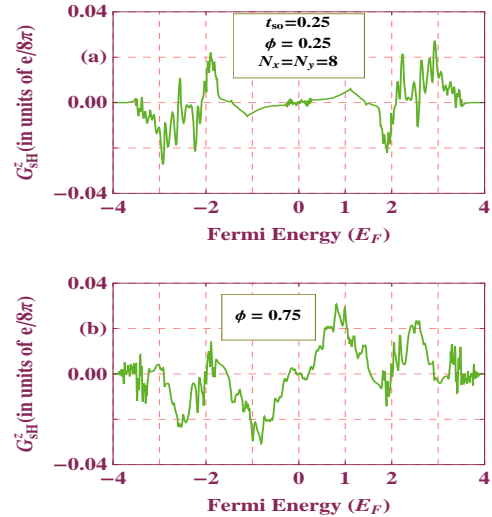}}\par}
\caption{(Color online). Variation of $G_{sH}^{z}$ as a function of Fermi
energy for (a) $\phi=\frac{1}{4}$ (b) $\phi=\frac{3}{4}$.}
\label{hallmag1}
\end{figure}
band-center (see Fig.~\ref{hallmag1}). The application of an external 
magnetic flux breaks the time reversal symmetry (TRS), but still SHE is 
still observed though reduced due to the elastic scattering. 

The reason is that in the presence of an external magnetic field along 
the $Z$ direction, an ordinary Hall voltage appears along the 
transverse edges of the sample and since the transverse leads are voltage 
probes no charge current will flow through them. Therefore the tendency of 
accumulation of charges of both spins somehow tries to reduce the features 
of SHE, which is entirely an effect due to the spin accumulation along the 
transverse edges and the flow of spin current in that direction, if leads 
are connected.

Before we end this section, we would like to point out 
that in a kagome lattice geometry Nori {\em et al.}~\cite{nori6} have 
described several physical phenomena in the presence of a transverse
magnetic field. They have analyzed the formation of cusps in the 
variation of superconducting transition temperature as a function of 
magnetic field in this geometry at $\phi=\frac{1}{8}, \frac{1}{4}, 
\frac{3}{8}, \frac{3}{4}, \frac{7}{8}$, 
etc., considering lower order approximation in the lattice path integral 
method which directly reflects the lattice topology on the electronic
properties. Though our present analysis is quite different from their
analysis, but here we also establish that the SHE exactly reappears at
these typical values of magnetic flux $\phi$ which reflects the nature 
of the lattice model.

\section{Closing remarks}

In conclusion, in the present paper we have studied different aspects of 
mesoscopic spin Hall effect induced by Rashba type SO interaction in a 
kagome lattice geometry attached to four finite width probes both in the 
presence and absence of external magnetic flux. We have evaluated spin 
Hall conductance (SHC) and longitudinal conductance for a finite size system 
in the clean limit using four-terminal Landauer-B\"{u}ttiker formalism 
and Green's function technique. In the absence of magnetic flux, we have 
observed that due to the change in Fermi surface topology SHC changes 
its sign at certain values of Fermi energy, along with the band center. 
Unlike the infinite system (where SHC is a universal constant 
$\pm \frac{e}{8\pi}$), here SHC depends on the external parameters like SO 
coupling strength, Fermi energy, etc. We have shown that in presence of 
arbitrary magnetic flux, periodicity of the system is lost and the 
features of SHC get suppressed because of the weak elastic scattering, but 
not lost completely. On the other hand, at some typical values of 
flux ($\phi = \frac{1}{2}$, $\frac{1}{4}$, $\frac{3}{4} \ldots$, etc.) 
the system retains its periodicity again depending on the system size and 
the features of spin Hall effect (SHE) re-appears. 

It is also important to note that in our theoretical model we have included 
the effect of magnetic field through the phase factor in the hopping term 
and ignored the Zeeman term in the Hamiltonian since in this geometry, the 
Zeeman coupling strength ($\sim 0.03$\;meV) is much smaller than the hopping 
integral ($\sim 1.5$\;meV) which has been discussed in the earlier
work~\cite{kimura}.

\end{document}